\documentclass[useAMS,usenatbib,onecolumn]{mn2e}
\usepackage{graphicx}

\title[Sun's retrograde motion and violation of even--odd cycle rule in 
sunspot activity]
{Sun's retrograde motion and violation of even--odd cycle rule in 
sunspot activity}

\author[J.Javaraiah]{J. Javaraiah\thanks{E-mail: jj@astro.ucla.edu}
\thanks{Permanent address :
 Indian Institute of Astrophysics, Bangalore-560 034, India
(Email address: jj@iiap.res.in)}\\
Department of Physics and Astronomy, 430 Portola Plaza,
UCLA, Los Angeles, CA 90095, U. S. A.}
\begin{document}
%\date{.....}

\pagerange{\pageref{firstpage}--\pageref{lastpage}} \pubyear{2002}

\maketitle

\label{firstpage}

\begin{abstract}
 The sum of sunspots number over an  
odd numbered 11 yr sunspot cycle exceeds that of its preceding 
even numbered cycle, and it is well 
 known as Gnevyshev and Ohl rule 
(or G--O rule) after the names of the authors who discovered it in 1948. 
 The G--O rule can be used to predict the sum of sunspot numbers of a 
forthcoming odd cycle 
from that of its preceding even cycle. But this is not always possible  
 because occasionally the G--O rule is violated. 
 So far   no  plausible 
 reason is known either for the G--O rule or the violation of this rule.
 Here we showed  the epochs of  the violation of the 
G--O rule are close to  the epochs of  the Sun's 
retrograde orbital motion about  the centre of mass of the 
solar system (i.e., the epochs at which  the orbital angular momentum 
of the Sun is weakly negative).
Using this result  easy  
to predict the 
 epochs of violation of the G--O 
rule well in advance.  We also showed that  
the solar  equatorial rotation rate determined from 
sunspot group data during the period 1879--2004 is  correlated/anti-correlated 
to the Sun's orbital torque during before/after 1945. We have 
 found the existence 
of a statistically significant $\sim$ 17 yr periodicity in 
the solar equatorial rotation rate. 
     The implications of these findings for understanding the  
 mechanism behind the  solar cycle  and the solar-terrestrial
 relationship  are discussed.

\end{abstract}


\section{Introduction}
Solar activity vary on many time scales.  It 
 can impact climate and the near--earth space environment 
\citep[e.g.][]{hs97,het99,roz01,hm04,geet05}. 
Therefore, prediction of the amplitudes of the variations in solar activity
 will  greatly help the  society.  
 Sunspots are the earliest observed phenomenon of solar activity. 
The sunspot cycles are numbered from the cycle 
which was began in the 
year 1755 (cycle~1). The current sunspot cycle,  which  began in the 
year 1996, is an odd numbered cycle (cycle~23).  
The well known Gnevyshev--Ohl rule or G--O rule \citep{go48} states
 that the sum of sunspot numbers ($R_{sum}$)  over an odd numbered sunspot 
cycle exceeds that of 
its preceding even numbered sunspot cycle. 
By using the G--O rule, it is possible to predict the $R_{sum}$ of an 
odd numbered cycle from that of its preceding even numbered cycle  with a 
reasonable accuracy \citep{wil88}. However, some pairs of the even--odd 
cycles violated  the G--O rule, i.e., in such a pair the $R_{sum}$ of the 
odd number cycle is less than that of  its preceding even numbered cycle 
(e.g., cycles' pairs 4,5 and 22,23).
 So far no  plausible  reason is known either for the  
G--O rule or its violation. In order to predict 
the amplitude of an odd cycle by using the G--O rule 
  it is necessary and essential  to know 
in advance whether the  even--odd numbered cycles' pair 
will satisfy    the G--O rule or not. But there is no method  available 
  for predicting the violation of the G--O rule.   
Predictions  on the basis of precursor technique, 
the G--O rule, and the statistical analysis of preceding cycles indicated a 
high $R_{sum}$ for the current cycle~23, similar to or exceeding that 
in cycle~22 \citep{jet97}. 
 The prediction of the  
 violation of the G--O rule by the cycles' pair 22,23  based  on the  
 long--term trends  in  sunspot activity \citep{sch55,kb01,jj03b}
 seems to be right. 
 However,   the epoch
of the next violation of the G--O rule  
 is not yet predicted,  and  
the available sunspot data may be inadequate to 
use this method.  

There are two main approaches for explaining the mechanism of the solar 
cycle, viz., one is based on a turbulent dynamo operating in or 
immediately below the 
solar convection envelope  and the other is a large scale oscillation 
superposed on a fossil magnetic field in the radiative core.
According 
to the turbulent dynamo theory the solar differential rotation
 produces a 
toroidal field (east--west component)
 by continuously winding up a poloidal field 
(north--south component),  induction effect of cyclonic turbulence regenerates 
the poloidal field, and the excess poloidal and toroidal fields are 
 removed by  the enhancement of 
diffusion by convective turbulence.  
A sufficiently 
detailed and realistic model of the dynamo process to account for all the
 different aspects of the solar magnetism is not yet available. The available
 turbulent dynamo models have several difficulties. For example, in these  
 models the role of the differential rotation in the cyclic variation of the
 solar activity is not clear;  
the reason for the cycle--to--cycle modulations of solar activity
   is not yet found,  and have no predictive power. 
The basic idea of the magnetic oscillator models is to consider the observed 
oscillating large--scale solar magnetic field as effects of 
periodic amplification of the primordial fields due to oscillations in the 
differential rotation rate of the solar interior. 
The main difficulty 
in the oscillator models is regarding energetics. No oscillator model offers 
the means of maintaining the oscillations against dissipation of velocity 
and magnetic fields \citep[see reviews by][]{rw92,os03}. In this regard  
it may be worthwhile to investigate whether 
the solar system dynamics could influence  on the internal dynamics of the 
Sun \citep{gj95}. 

The idea that the gravity of the planets might be the cause of the solar cycle 
dates back at least to Carrington \citep{brown00}. 
 Subsequently,  
  many scientists suggested a possibility of     
 the tidal forces due to  planets 
or the rate of change of the Sun's orbital angular momentum 
 about the centre of the solar 
system (barycentre)  having a role in the mechanism of  solar activity.
Such an idea of a role of the solar system dynamics has been doubted 
because, \citep[see][]{fe69}:  (1) the energy of the  tidal force due to the planets
 is small compare to 
the  Sun's surface gravity; and (2) the Sun's 
centre of mass is free--fall  in the sum--total gravitational 
field of all the planets.
Nevertheless, 
the hypothesis of a relationship between the Sun's motion  
 and the solar activity is 
supported by a growing number of studies
indicating 
something  must be true in the `planetary hypothesis'
\citep{jo65,ww65,bl83,bl89,fs87,sf90,gok96,zaq97,la99,ch00,ju00,ju03}.

The Sun wobble about the centre of the solar system 
with the distance varying up to 2 times its radius. 
The Sun's spin momentum contributes 1--2\% to the
 total angular momentum of the 
solar system.  
\citet{jo65} showed  the existence of a relationship between 
 a Hale cycle  and  
  the changes in the 
 angular momentum of the Sun's motion  about 
the barycentre. 
Recently,  
\cite{zaq97} and \cite{ju00} found that the Sun's motion 
about the barycentre is 
having a role  even in  
 the cause of the solar differential rotation.  
The configurations and the directions of 
alignments of the major planets are considerably different 
during the even and the odd numbered cycles \citep{ms79}.
The differential rotation 
analysis of  \citet{jg95}
 and \citet{jj96,jj03a}  revealed the  
frequencies  which  are 
compatible with the frequencies of the specific alignments of two or more 
planets.
 The existence of a relationship,  which is similar to     
 the G--O rule in sunspot activity, is also found 
 between the differences in the differential 
rotation during the odd and the even numbered cycles \citep{jjet05a}.
Therefore, one would reasonably expect that  
the violation of the G--O rule 
and the variations in the differential rotation 
 are probably  having a relationship with 
  the Sun's motion about the barycentre.
In the present paper we have investigated this. 

In the next section we describe the data and the analysis. 
In Section~3 we show 
the existence of a relationship between 
 the violation of the G-O rule in sunspot activity  
and the Sun's retrograde motion about the centre of mass of the
 solar system. In Section~4 we show the existence of 
coupling in the Sun's 
spin and  orbital motion. In Section~5 we discuss about 
the implication of these results for understanding the   
long--term variations in the solar activity 
 (including the Maunder minimum) and the solar-terrestrial 
relationship.

\section{Data and analysis}

Dr. Ferenc Varadi kindly provided us the values of      
the following :  the distance (R) of  the Sun's centre 
from the solar system barycentre,  the orbital  velocity of
 the Sun (V), 
 the orbital angular momentum of the Sun (L), and the rate of change of the 
orbital angular momentum (orbital torque  $\frac{dL}{dt}$), for each 
 interval of 
length  10 days during the period 1600--2099. 
He determined these values using the recent  
   Jet Propulsion 
Laboratory (JPL) DE405 
ephemeris \citep{sei92,stand98} for the period 1600--2100. 

The solar differential rotation can be determined
from the full disk velocity data using the standard polynomial 
expansion \citep{hh70}, 
$$\omega(\phi) = A + B \sin^2 \phi + C \sin^4 \phi , \eqno(1) $$
while for sunspot data it is sufficient to use only the first two terms 
of the expansion \citep{nn51} :
$$\omega(\phi) = A + B \sin^2 \phi ,  \eqno(2)$$
 where 
$\omega(\phi)$ is the solar sidereal angular velocity at 
latitude $\phi$, the coefficients  $A$  represents
the equatorial rotation rate and $B$ and $C$  measure the 
latitudinal gradient in the rotation rate 
with $B$ representing mainly low latitudes and $C$ largely
higher latitudes ($C$ is too small to be determined from sunspot data).   
(Note: The above equations  have no theoretical  foundation, but  
 fit very well to the corresponding data, said above.)

In this analysis the sunspot data  and its reduction are  same as 
in  
\citet{jg95} and \citet{jj03a,jj03b}. 
We have used   the Greenwich  data on sunspot 
groups during the period 1879--1976    
and the spot group data from the Solar Optical Observing Network (SOON) 
 during the period 1977--2004 
(available at ftp://ftp.science.mfsc.nasa.gov/ssl/pad/solar/greenwich\break.htm). 
The data consist of the observation time, heliographic latitude and longitude, 
central meridian distance (CMD), etc.,  
for each spot group on each day of its observation. 
 The sidereal rotation velocities ($\omega$) have been computed for 
each pair of consecutive days in the life of each spot group using 
its longitudinal 
and temporal differences between these days.  
 We have not used the data
 corresponding to the $|CMD|>75$ deg
on any day of the spot group life span and the displacements exceeding
3 deg in longitude or 2 deg in the latitude per day. 
We  determined the annual 
variations in the coefficients $A$ and $B$ by fitting  
the each year spot group data  to the 
equation~(2).

\section{Violation of the G--O rule}

Fig.~1 shows 
 the variations in the R, V, L, and $\frac{dL}{dt}$
 determined from the planetary 
 data available for every 
  10 days,   
 during the period 1600--2099.
In this figure we also show  the variations in the yearly mean values of 
the sunspot numbers during the period 1600--2004 (bottom penal).
 The  epochs 1632, 1811 and 
1990 when  the Sun's motion about the barycentre was retrograde  
\citep[i.e.,  when L was changed from positive 
to a weakly negative,][]{jo65} are indicated by the dotted vertical lines.
The other two epochs of the big drops in L were at  1672 and  1851, 
 and the expected next epoch of  such a big drop in L will be at 2030. 
All these are indicated by 
the dashed vertical lines.   
 For obvious reason near each of the big drops in L there is a big drop  
of $\frac{dL}{dt}$.   
We  have given in Table~1 the values of R, L, V, $\frac{dL}{dt}$, 
and the values of the ecliptic longitudinal positions 
 of the giant planets at 
these 6 epochs. 
 The phase of $\frac{dL}{dt}$ is leading  the phase of L by about 4 years near 
the dotted lines and about 5 years near  the dashed lines.
 This is expected because the former is the force and the latter 
is the motion and both have main period, 19.86 yr, the conjunction period 
of Jupiter and Saturn. 
At the epochs where the steep decreases in L are 
indicated by the dotted vertical 
lines the decrease in R is more steeper than the decrease in V. 
 It is opposite in  case of  
 at the epochs which  are marked by the 
dashed vertical lines. 
 The gap between the consecutive dotted lines and also between 
the consecutive   
dashed lines is about 179 years, i.e., the period of the well known 
 179 yr cycle in the Sun's motion related to the
solar system barycentre. 
It is also the period of an interval between alignments of all the outer 
planets in a same configuration and in a same direction in space.
It is approximately equal to the 9 conjunctions periods of  Jupiter 
and  Saturn \citep{jo65}. 
The gap between a dotted line and its 
neighbour dashed line is about 43 years, the conjunction period of Saturn and 
Uranus.

\begin{table}
\centering
 \caption[]{The  values of $\frac{dL}{dt}$, L, R, V and the ecliptic  
positions (in deg) of the giant planets--Jupiter (J), Saturn (S), 
Uranus (U), and 
Neptune (N)--at the epochs for which the values of 
 L are given and  marked by the dotted and dashed vertical lines in Fig.~1. 
 The units of R, V, L and $\frac{dL}{dt}$ are
$A.U.$, 
 $A.U.$ day$^{-1}$, 
$M_\odot \times (A.U.)^2$ day$^{-1}$ and 
$M_\odot \times (A.U.)^2$ day$^{-2}$,  
 respectively, 
 where $M_\odot$ is mass of the Sun and $A.U.$ is the Astronomical unit.} 

\begin{tabular}{lcccccccccc}
\noalign{\smallskip}
\hline
\noalign{\smallskip}
Time & $\frac{dL}{dt}$ $\times 10^{-9}$&& Time & L $\times 10^{-8}$ & R$\times 10^{-3}$& 
 V $\times 10^{-6}$&  J & S & U & N \\
\noalign{\smallskip}
\hline
\noalign{\smallskip}
1628.64 & -2.34&& 1632.28 &0.039 & 0.747 & 5.773 & 37  & -120 & 178 &  -138 \\
1666.97 & -2.53&& 1671.79 &0.153 & 0.303 & 5.094 & 157 & -8   & -13 &  -53 \\
1807.64 & -2.40&& 1812.05 &0.033 & 1.334 & 5.535 & 96  & -87  & -127 & -106 \\
1845.97 & -2.51&& 1850.85 &0.227 & 0.462 & 4.941 & -171 & 20  &  30  & -22  \\
1986.81 & -2.62&& 1990.97 &0.021 & 1.364 & 5.285 & 126  &-63  & -80  & -75 \\
2025.14 & -2.28&& 2029.99 &0.310 & 0.638 & 4.940 & -138  & 53  &  76  &  10 \\
  \noalign{\smallskip}
\hline
\end{tabular}
\end{table}

The G--O rule was violated by the 
 sunspot cycles' pair 4,5 at the beginning of the Dalton minimum,  
  and it is
 most likely  to be violated by  the cycles' 
pair 22,23 \citep[e.g.][]{jj03b}. 
It seems that
near the end of the cycle 
which was
at just one cycle 
 before the cycle at before the beginning of the Maunder minimum 
  the G--O rule  
 was violated (cycles' pair -12, -11, say).  
 \citep[have shown   the existence of  cyclic behaviour
 during the Maunder minimum.]{beet90}
 Interestingly, each of 
these cycles' pairs are   
close to an  epoch at which  
 the Sun's orbital  motion is retrograde, which are  
 indicated by the dotted vertical lines. 
 The
peak value of sunspot cycle~8 is higher than that of cycle 9. 
 By virtue of 
this difference the cycle pair 8,9  violated the G--O rule.
The temporal behaviours of L and $\frac{dL}{dt}$ suggest such a 
situation might have occurred in the year   
1672 and the next such a situation may  occur near the year  2030.
These findings indicate the existence of   
  a relationship between the violation of the G--O rule  
 and the Sun's retrograde motion about the centre of mass of the 
solar system.
In Table~1 it can be seen that  the epochs  at which   L was steeply decreased 
 Saturn was aligned  approximately 
in opposition to Jupiter, and Uranus  and 
Neptune were nearer 
to  Saturn 
 (i.e.,  Jupiter leads by about 180 deg w.r.t. 
the other three giant planets). 
Obviously such  configurations  of the major planets are responsible 
for the Sun's retrograde  motion about  the  barycentre, which is  in turn    
seems to be responsible for  the violation of the G--O rule. 
Since the planetary configurations  and  
the Sun's retrograde  motion 
can be computed well in advance,  hence,  
it is  possible to know the epochs of violations of the G--O rule  
well in advance.  Therefore, the G--O rule is expected to be violated   
 by the  Hale cycle  which will include (or end at)  
the year 2169, i.e., only after 
a gap of about eight Hale cycles after the current Hale cycle~11. 
However, the violation by virtue of the difference in the 
 heights of the peaks of the cycles 
--like  the cycles' 
pair 8,9 near the year 1851--is expected  to be happening near the year 2030,
 i.e., by  the
 cycles' pair 26,27.

\clearpage
\begin{figure}
\centering
\includegraphics [width=14.0cm]{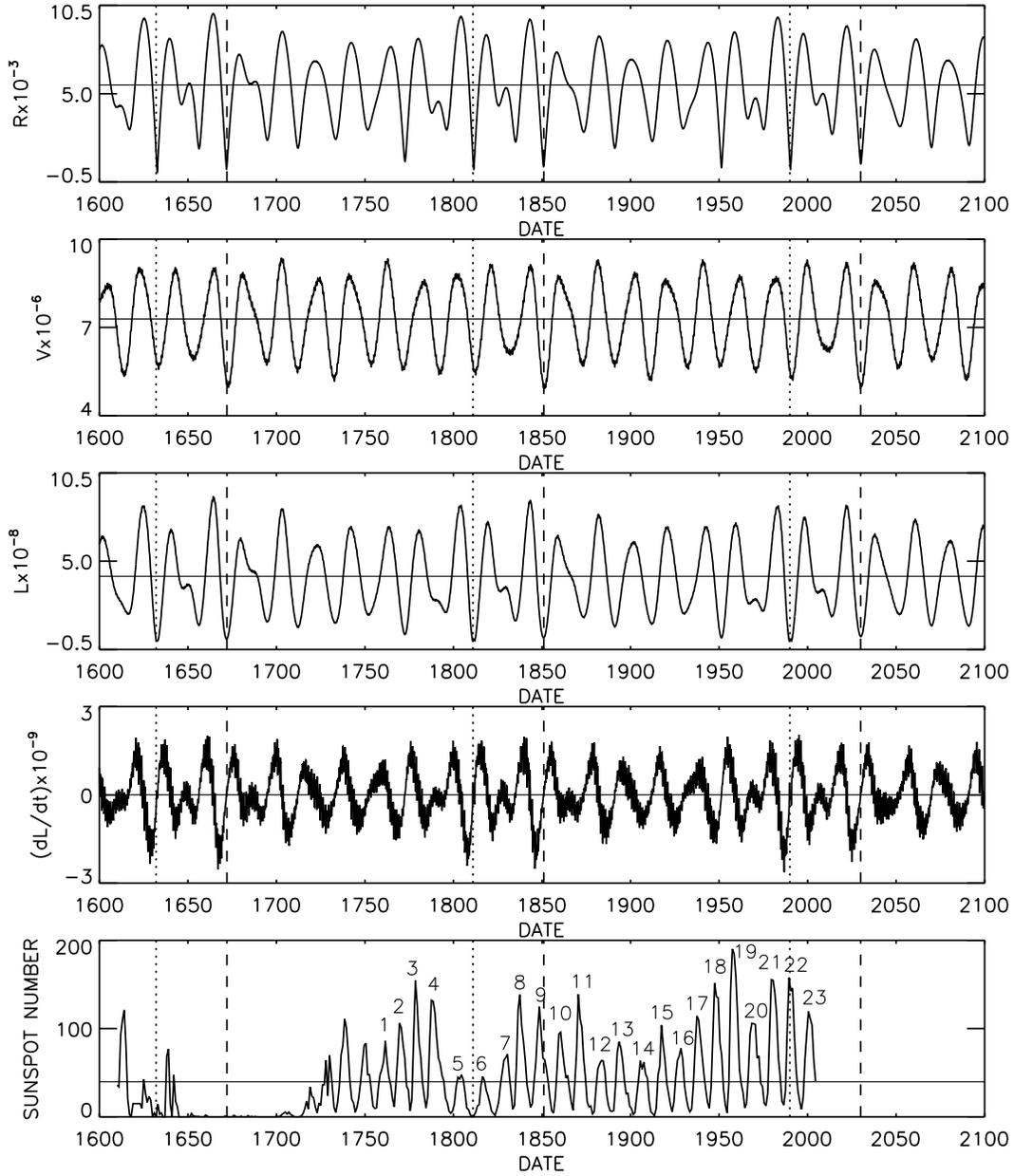}
\caption{Values of R, V, L, $\frac{dL}{dt}$ in each 10 days interval
 during the period 1600--2099 and the annual mean of the 
number of sunspots  during the period 1600--2004.
The units of R, V, L and $\frac{dL}{dt}$ are 
 $A.U.$, 
 $A.U.$ day$^{-1}$, 
$M_\odot \times (A.U.)^2$ day$^{-1}$ and 
$M_\odot \times (A.U.)^2$ day$^{-2}$,  
respectively, 
 where $M_\odot$ is the mass of the Sun and $A.U.$ is the astronomical unit.
 Near the peak of each sunspot cycle the corresponding Waldmeir 
sunspot cycle number is  marked. 
The epochs, 1632, 1811 and 1990, at which the Sun's orbital 
motion was retrograde are indicated by the dotted--vertical lines. The other 
three epochs, 1672, 1851 and 2030, where  also L is steeply decreased 
 are marked by the dashed--vertical lines. The horizontal lines 
represent the mean values.}
\end{figure}
\clearpage

\section{Sun's spin--orbit coupling}

Fig.~2 shows variations in the annual average values of 
 R, V, L, $\frac{dL}{dt}$ and the  values of 
the differential rotation parameters
 $A$ and $B$ determined from the yearly sunspot group 
data during the period 1879--2004.
The error bars are $\pm$1$\sigma$ (standard deviation) values.
Due to the reduced number of sunspot groups the values have large errors  at 
 the cycles' minima.
 We corrected the 
time series of $A$ and $B$ by replacing  
 the values which have error larger than three
times the median errors by the  values simulated from the linear fits 
\citep[a similar correction was applied in an earlier paper by][]{jk99}.
 In  Fig.~2 the solid curves in the lower two panels 
 connected the points  of the corrected data and the dotted curves connected 
the uncorrected  
 data.   In this figure  the variations in the solar 
equatorial rotation rate, $A$,  looks  to be largely similar to 
the variation of $\frac{dL}{dt}$. After 1945  the variations of 
 both $A$ and $\frac{dL}{dt}$ have  
 somewhat large amplitudes. During this time  the mean level of activity 
is also relatively large (see Fig.~1).  
The epoch, 1990--1991,  at which the Sun's orbital motion is retrograde 
the value of $A$ is low and the $\frac{dL}{dt}$ is almost zero.
The correlation between $A$ and $\frac{dL}{dt}$ is positive  before around 1945 and 
negative  after that time (Corr. Coeff., $r =  40$ and $-50$ in intervals of about 
50 years before  and after 1945, respectively). 
 These results indicate the existence of a relationship between $A$ 
and $\frac{dL}{dt}$.
The orbital angular momentum might have  been transfered to the spin momentum 
for about 50 years before near 1945  and the reverse might have been 
happened in the
 latter 50 years \citep{ju00}.
    
The correlation between the latitudinal 
gradient of the rotation ($B$)  and $\frac{dL}{dt}$ is weak.
The signs of correlations ($r\approx$ -20 to +25) between the 
 ($B$) and $\frac{dL}{dt}$
  are found to be   opposite  
 during the  aforesaid epochs of the positive and the 
negative  correlations between  $A$ and $\frac{dL}{dt}$ 
 (it should be noted here 
that   there exists 
an anti-correlation  between $A$ and $B$, e.g., Javaraiah et al. 2005a). 

\citet{jg95} and \citet{jk99} found 
the $\sim$18.3-yr, $\sim$8-yr and few other short periodicities in $B$.  
Fig.~3 shows the  FFT spectra of the annual variations of
 $\frac{dL}{dt}$, $A$ and $B$. From this figure it can be seen that  
 both spectra of $A$ and $B$  have the dominant  
 peaks at frequency 1/18 yr$^{-1}$, which  
 are significant on $3.6\sigma$ and 5.5$\sigma$ levels
respectively. The corresponding 
 periodicities in $A$ and $B$ match approximately with that of 
the main periodicity in 
   $\frac{dL}{dt}$ (the peak at 1/21 yr$^{-1}$, 6.6$\sigma$). 
 We  repeated the FFT analysis by extending the time series
 from the original 
126 data points to 1024 data points  by padding the time series 
with zeros. The values of the aforesaid main periodicities in $A$, 
$B$ and $\frac{dL}{dt}$ are 
 found to be 17.1 yr, 18.29 yr, 19.69 yr, respectively.  

Note there is about 1 yr difference between the aforesaid main periodicities
 of $A$ and $B$. This may be explained as follows:
 It is believed that the magnetic structures of the active regions 
originate  near the base of the convection zone (about 200 000 km below 
the surface) and the magnetic bouncy cause  them to rise through 
the convection zone and emerge on  
   the surface.  
The rotation rates of spot groups 
depend on their life spans and  age. This can be interpreted as the 
 rotation rates of the magnetic structures of spot groups  
vary as their anchoring depths vary during their life spans 
\citep{jg97,gj02,hi02,siet03}. 
Fig.~4 shows the FFT spectra of $A$ and $B$ determined 
from  the first two days data (young groups)
 of the spot groups of life span 7--12 days.
  To have adequate data we have 
used the  moving time intervals of sizes 5 years  
\citep[same as in][]{jg95,jj98}.
In Fig.~4 we have also showed the FFT  spectrum determined from
 the 5-yr smoothed time series of $\frac{dL}{dt}$. 
In this figure it can seen that the dominant peaks in the spectra of $A$ 
and $B$ are well coincided.
We  also repeated the FFT analysis for these smoothed time series 
by extending them as described above.   The dominant peaks 
 in the spectra of 
the extended time series of both $A$ and $B$  are found to be  
at 1/17.4 yr$^{-1}$. This  indicates that the 
 18.3-yr periodicity--found in $B$ derived from the combined data 
(dominated by small and short lived groups)  of 
the spot groups of different life spans   and age--may 
 correspond to a slightly shallower layers.  The 
 17.1 yr periodicity may correspond to a 
slightly deeper layers \citep[also see][]{jj98}. 

The 17.1 yr  periodicities of both $A$ and $B$   closely match with the 17.5 yr 
period found in the mixing of the low frequency components of L and 
the  instantaneous spin projection vector \citep{ju00}. 
In addition, it seems that there exists a good agreement between 
the  amplitudes of the variations in the Sun's spin and the orbital 
angular momenta, particularly at the common epochs of the steep decreases
 in both
L and $A$. At these epochs the amount of decrease in L is approximately 
equal to the mean value of L. For example, 
at  the epoch 1990.97 the amount of the steep decrease in L is about  
 $-2.1\times 10^{47}$
 g cm$^2$ s$^{-1}$. At this epoch 
 the amount of the drop  in $A$ is about 1\% and the corresponding 
decrease in the spin momentum is found to be approximately  
  $-1.1\times 10^{47}$
 g cm$^2$ s$^{-1}$. 
\citep[The mean yearly value of $A$ is
 14.505 $\pm$ 0.008 deg day$^{-1}$. The uncertainty, 1$\sigma$ value, 
in this mean 
value  
suggests that the mean amplitude of the yearly variation in the 
solar equatorial 
rotation rate during the period
 1879--2004 is about 0.056\% only. 
Overall about 0.1\% difference is found  between
 the mean equatorial rotation rates
during the even and  the odd cycles,][]{jj03a}.   

The  results above   
provide a direct observational support to the models of   
the spin--orbit coupling of an oblate Sun  \citep[e.g.][]{ju00}. The results 
also indicate that the 
 perturbations required for maintaining the oscillations  
in the solar differential rotation and the solar magnetic field 
as the  participants in the mechanism of solar cycle 
 are coming from the solar system dynamics. 

\clearpage
\begin{figure}
\centering
\includegraphics[width=14.0cm]{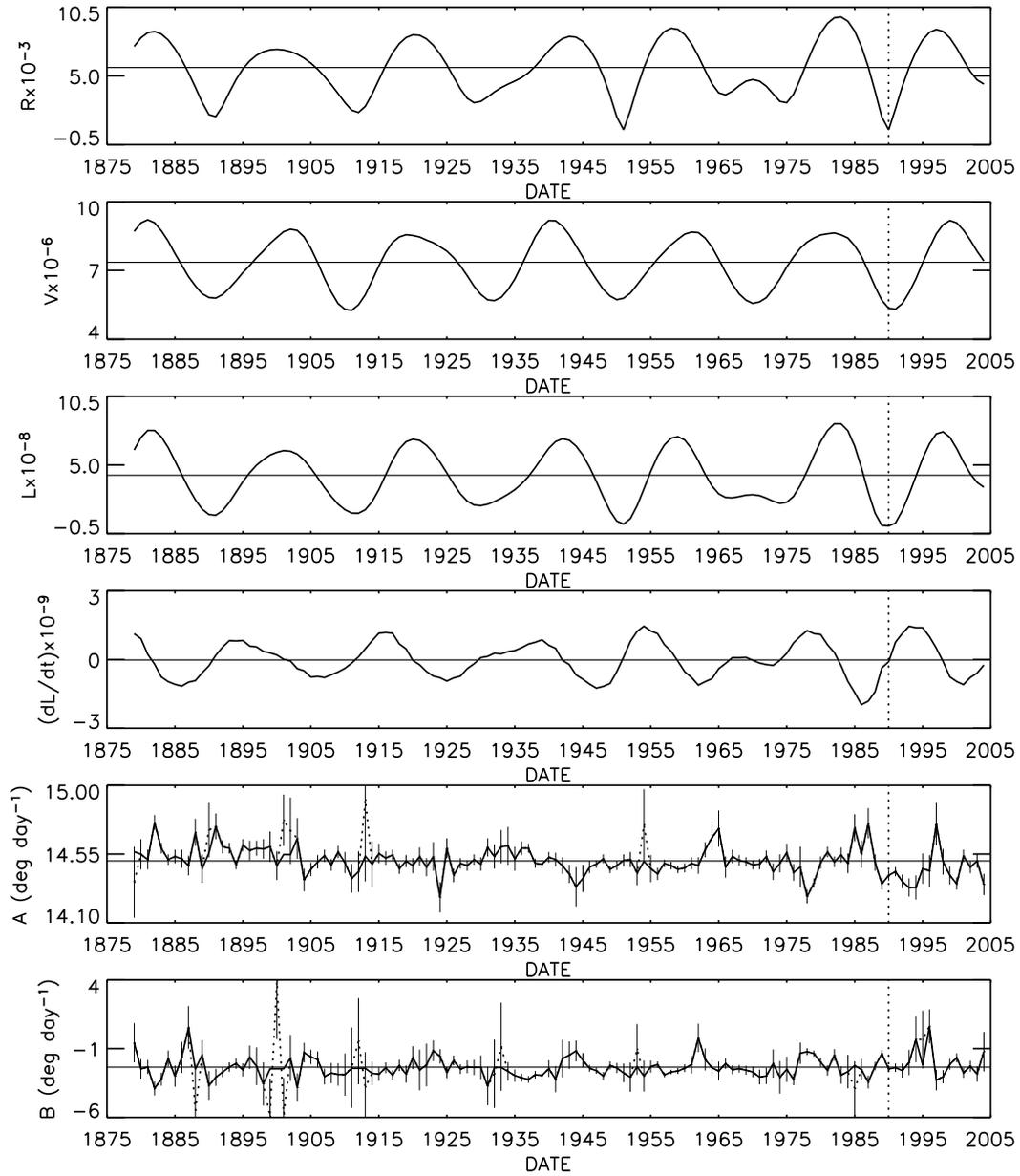}
\caption{Variations in the yearly mean values of R, V, 
L, $\frac{dL}{dt}$, $A$  and $B$.
The units of R, V, L and $\frac{dL}{dt}$ are the same as  
 in Fig.~1 and Table~1.  
In case of $A$ and $B$  the solid and dashed curves 
represent the corrected  and  the uncorrected  
  data, respectively, and the 
error bars are 1$\sigma$ 
 values. The epoch 1990 at which the orbital motion of the Sun 
was retrograde is indicated by the dotted--vertical line.
 The horizontal lines represent the mean values.} 
\end{figure}
\clearpage

\begin{figure}
\centering
\includegraphics [width=14.0cm]{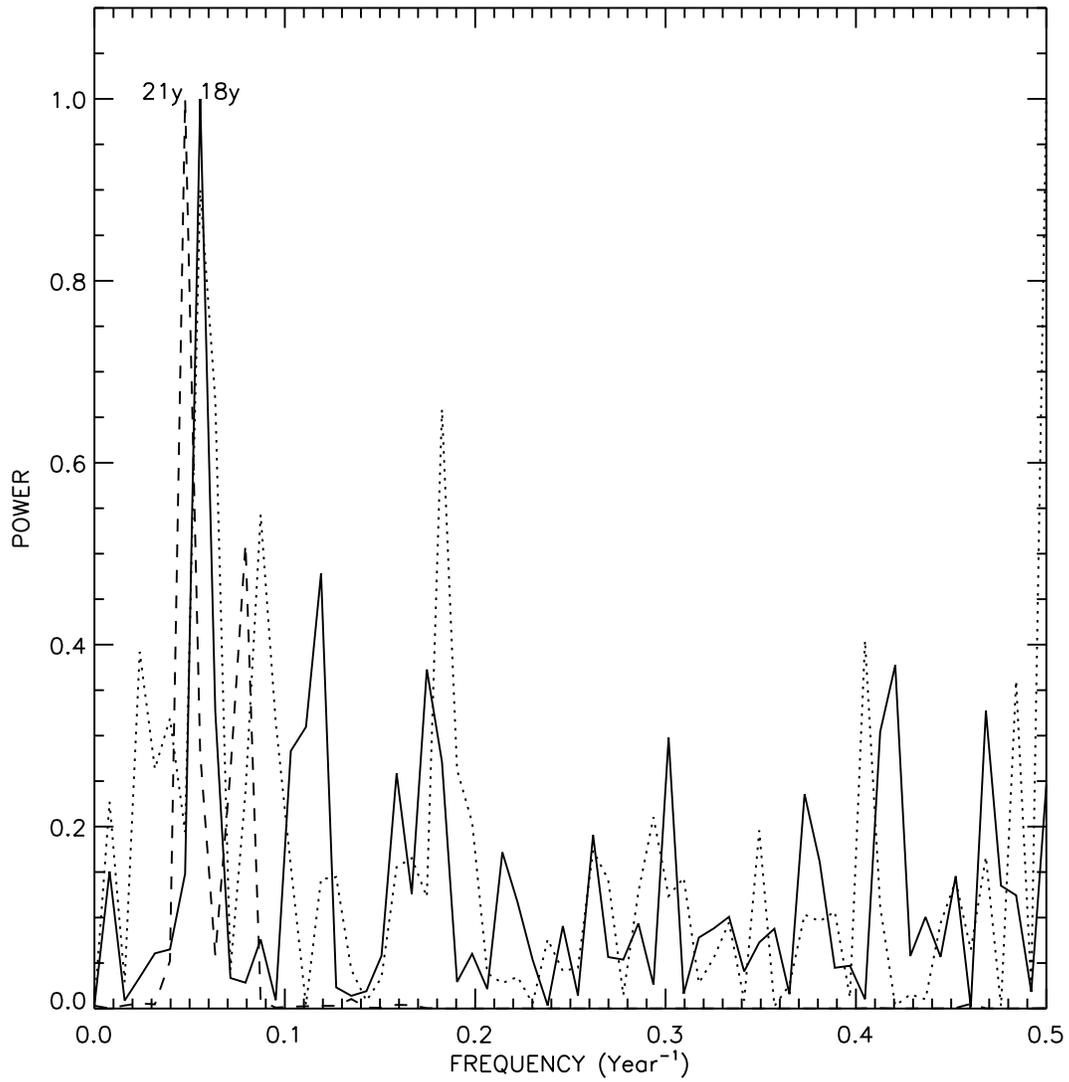}
\caption{FFT power spectra of  $\frac{dL}{dt}$ (dashed curve),
 $A$ (dotted curve) 
and  $B$ (solid curve). The power values are normalized to unity. 
Near the tops of the dominant peaks, which are significant on $>\ 3\sigma$ 
(particularly in $A$ and $B$), 
the values of the corresponding  periods are shown.}
\end{figure}
\clearpage

\begin{figure}
\centering
\includegraphics [width=14.0cm]{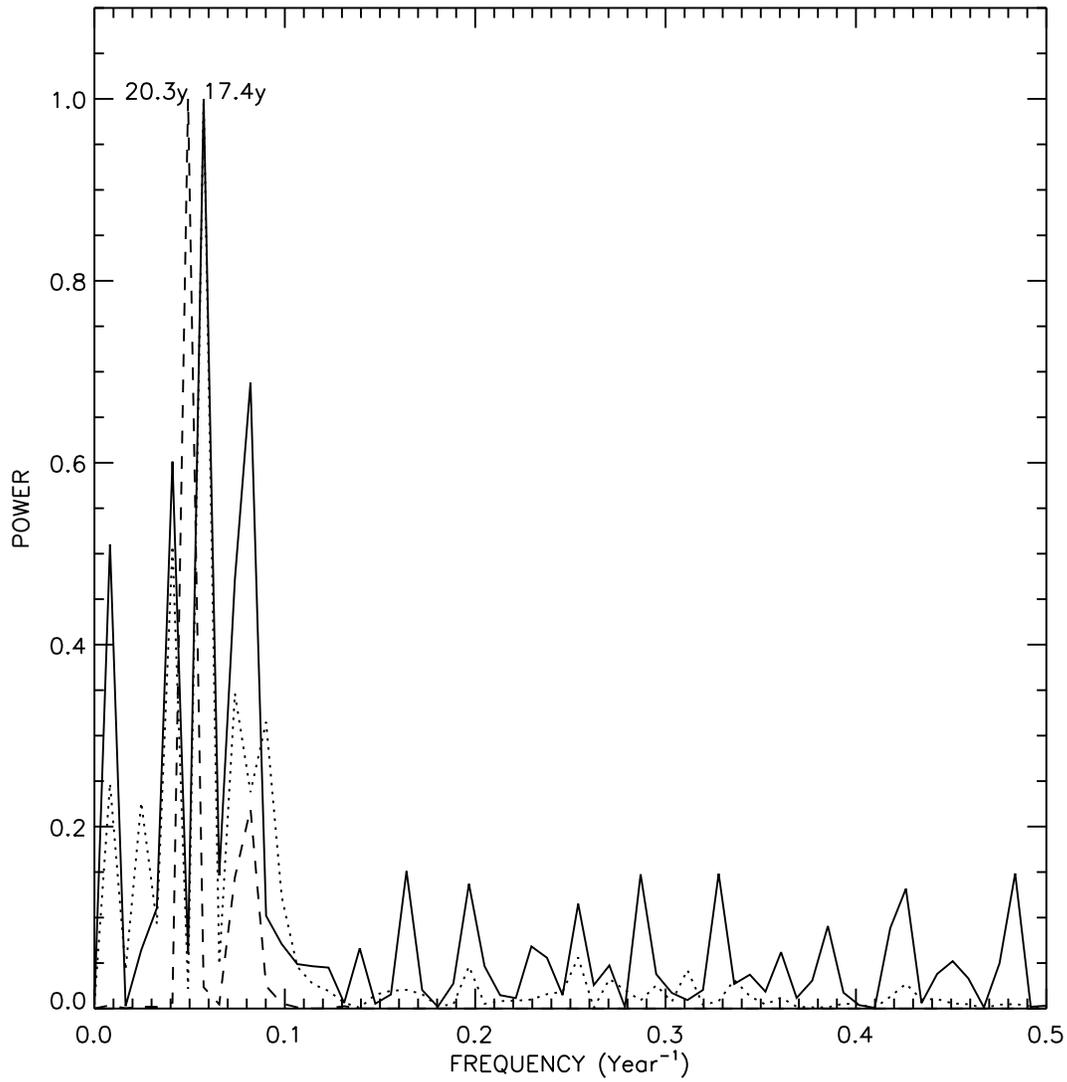}
\caption{Same as Fig.~3, but $A$  and  $B$ are determined from only 
first two days (young groups) data of the spot groups of life span 7--12 days. 
To have adequate spot group data 5-yr moving time intervals are used. 
In case of $\frac{dL}{dt}$, 5-yr smoothed time series is used.}
\end{figure}
\clearpage

\section{Discussion}
\citet{uset01}
 argued that  
 between sunspot cycles~4 and 5 the data is sparse and  unreliable,  
 and interpreted the very  
long  cycle~4 as consisting of two short cycles. If this interpretation 
is correct, then it seems that the G--O rule was not violated by  Hale 
cycle~2. \cite{kriet02}   
  by comparing  
the sunspot observations of the aforesaid period 
with those at other times, and also by analysing other proxies of 
solar activity, showed  that no cycle was missed at 
the end of the 18th century and that 
the official sunspot cycle numbering and parameters are correct.     
 \citet{uset03} argued     
 the statistical analysis performed 
in the paper by Krivova et al. (2002) as it is not 
validated by quantitative tests 
and even contains several errors.
Hence, whether 
 the G--O rule was violated during the Hale cycle~2 or there was  
an additional weak cycle in 1790's, is yet to be confirmed.
The result found  in Section~3   
  essentially 
 strongly suggests that 
 the G--O rule was indeed violated by the Hale cycle~2.

  \citet{bl89} found  a depressed level of 
activity in a few cycles which follow the epochs of the 
retrograde orbital motion of 
the Sun. This can be seen in  
 Fig.~1, i.e.,  the level of activity is relatively   
  low during at least   a few cycles which follow
 the dotted vertical lines.      
 (Around the years 1730 and  1900 the level of  activity 
was considerably low, whereas the sizes of the drops in L and R were not 
 large, but 
there were considerably large drops  in V. In 1900 there was opposition 
alignment,    about 25 deg.,  of all  the 
other major planets with Neptune.)  
As already mentioned in Section~3 the expected violation of the G--O
rule close to the  
 dotted vertical line at 1632 is 
followed by the Maunder minimum, and the G--O rule violation
 close to the dotted vertical line at 1811 is  
followed by the Dalton minimum. 
Therefore,  violation of the G--O rule close to the
 dotted vertical line at  1990 is also expected to be  
followed by a Maunder/Dalton like minimum in activity. 
 That is, the present trend of the relatively   
low level of sunspot  activity in the current cycle~23, which follow the
dotted vertical line at 1990,  may  continue for a few more sunspot cycles. 
Such an indication is also found in the recent studies of the  
long--term variations in sunspot activity \citep[e.g.,][]{boet04,hw05} and 
 solar equatorial rotation rate \citep{jj03b,jjet05b}.
In addition,  a number of authors predicted 
a  weak  activity during the  next cycle~24,  
  using a large number of techniques
 \citep[e.g.,][]{kane99,echer04,sval05}.
Therefore, 
 violation of the G--O rule   may be  an indication of 
 onset of a Maunder/Dalton like minimum in activity.
All these results   seem to be consistent with the result,   
 the 179 yr cycles of coherence between the solar magnetic  
activity and the solar system 
dynamics, was  found by \cite{jo65}.  
If we consider that  the latest three consecutive 179 yr cycles in sunspot 
activity  
began near the years  
1632, 1811 and 1990,   
 the first half (may be represent a  Gleissberg cycle)  of 
each of the first two 179 yr cycles   is weaker than the 
corresponding  second half, 
suggesting 
that the current half 179 yr cycle (includes about 80\% of current century)
 would be weaker than the later half. 
Within this half 179 yr cycle  
 the average activity during the first half (the double Hale cycle which 
 comprises the cycles~22--25) may be 
weaker than that during  the second half 
\citep{jj03b}. 
 (Note  two major drops in L occur
 within the first quarter of a 179 yr cycle, 
 in a time gape of 
about 43 years.) In fact, the current epoch of sunspot activity 
seems to be at the declining phase of the Gleissberg cycle whose 
minimum is expected to be occurring near the end of cycle 25 \citep{jjet05b}.

The Maunder minimum and other such low activity
epochs were also explained on the basis of variations in the  
Sun's motion about the centre of mass of the solar system  
\citep{fs87,ch00,ju00}. 
 One of the reasons often quoted for rejecting  a role of the  solar system 
dynamics in the  mechanism of the
solar activity is that  sunspot activity was 
absent during the Maunder minimum but the  
planetary configurations were present \citep[e.g.][]{smy77}.
 This argument seems to be not valid because  
the interval between the alignments of all the outer 
planets in the same configuration and in the same direction in space 
is about 179 years, 
 and no  alignment of the major planets
 repeats exactly \citep[and also see Table~1]{jo65}.
A considerable information on the solar rotation rate during the Maunder 
minimum is available. 
 \cite{egt76} analysed the sunspot drawings  made by 
J. Hevelius during the  
period 1662--1664, i.e.,
just before began of the Maunder minimum 
  and found that  
the equatorial  rotation rate 
was about 4\% higher than the value during  
the modern time. 
\cite{abrw81} analysed the same data 
 and found that the equatorial rotation rate  
was same as that during the modern time. 
\cite{rnr93} analysed  a unique  
 collection of sunspot 
observations recorded at the Observatoire de Paris from 1660-1719,
 and  
found that the equatorial rotation was about 2\% lower than that during 
the modern time.  
Recently, \cite{vaet02} analysed the observations of a sunspot carried 
out by \citet{fstd1684} from 25 April 1684 to May 1684 (Julian date) 
 and found that during the deep  Maunder minimum (1666--1700) the 
rotation rate  near   the equator was 
about 5\% lower than that of the modern time. 
Overall these  results suggest existence of a large drop in the equatorial 
rotation rate during the Maunder minimum. 
The  large drop in the equatorial
rotation rate during the deep Maunder minimum  
might be related to the  
steep decreases in
 L and $\frac{dL}{dt}$ at 1632 and 1671 (see Fig~1),  and obviously to 
the configurations of  
the major planets at these epochs (see Table~1).  
 The  effect of
 the large drop in $A$ near the beginning of 
the Maunder minimum    
 might have been persisted throughout 
the Maunder minimum  and caused the nearly complete absence of 
activity \citep{jj03b}.
 It is  interesting to 
note  that the beginning of the well know Sp\"orer minimum (1450--1550) 
is approximately about 179 years ago from the year 1632. 
Therefore, the cause of the  Sp\"orer minimum  may be also same as that 
 of the Maunder minimum, as suggested above.

We have used here the data on 
the giant planets only. However,  the inner planets also may be  
 important because of their proximity to the Sun  
their tidal forces on the Sun are larger than those of the 
outer planets (except Jupiter).
Therefore,  when they
are  closely align with  
 the Jupiter the combined effect may cause `jerks' (rate of change of
 acceleration) in the  
orbital motion of the Sun \citep{ww65}. 
There are some spikes in the variations in $A$ and $B$, 
particularly around some minimum years the values of $B$ are 
 almost zero or even positive (see Fig.~2).
 Many of these spikes may be resulted because of the sizes 
of the spot group data are 
 small  during these years. 
In the variation of $B$ the spikes at the years 1887, 1962 
and 1996 have  
considerable influence on the significance levels of the periodicities in $B$ 
\citep{jk99}. But during these years 
the statistics are sufficiently good 
and the values have errors less than three times the median error. 
Moreover,  we found the similar abnormal behaviours in the values of $B$ during the
 1962 from   
  the Mt. Wilson Observatory and the Kodaikanal Observatory 
 sunspot data 
 (available at fft://ftp.ngdc.noaa.gov/STP/SOLAR$\_$DATA/SUNSPOT$\_$REGIONS/SUNSPOT$\_$REGION$\_$TILT/).  
 We confirmed the abnormal  behaviour of $B$   
during the year 1996   
using the data of sunspots drawings of Mt. Wilson Observatory 
(available at http://www.astro.ucla.edu/$\sim$obs/spotframe.html). 
\citet{kn90} also derived a similar value of  $B$ during the year
 1962 from the  
 spot group data measured in  the National Astronomical Observatory of Japan.  
Hence, the aforesaid abnormal behaviour 
of $B$ seems to be a real property of the rotations of the sunspots 
during the aforesaid years. 
Incidentally,  on
 February 5, 1962 
the five naked eye planets plus the Sun and Moon aligned within 15.8 deg 
and there was a solar eclipse at the same time \citep{mos96}.

In view of the existence of a statistical significant 
 $\sim$ 18.3 periodicity in $B$ it is interesting to note 
 that the  major droughts in the world \citep{hs97}
 and even the major 
earthquakes in California seem to be occurred in the gaps of about 18 years 
(http://www.personal.psu.edu/users/m/a/mab573/tsunami.htm). 
A similar periodicity my be exist in the Earth's rotation \citep{kiet02}.
The Precession period of the moon is also 18.6 years. 
 In addition, there were  sever droughts in 1886/1887, 
1962/1963 and 1995/1996 \citep{fow00}, when the values
 of the coefficient $B$
are abnormal. 
So, in view of the results in Sections~3 and 4, 
 it may be  worthwhile to investigate whether the variations in the  
internal dynamics of the Sun and the Earth, and  the terrestrial phenomena
all  governed by the solar system dynamics.  
 However, it should be noted here that so far no convincing evidence
 is found for the influence of  
the  planetary dynamics  on
 terrestrial phenomena, climate,  dynamics of the 
Earth and/or earthquakes. 
\citet{gp74} described the 1982 alignment of all the nine 
planets 
  as a super--conjunction with all the nine planets in a line on the same
 side of 
the Sun. They  had predicted that this alignment of planets cause a massive earthquake 
in 1982 and  a  
 major disaster in  Los Angeles. Fortunately, 
that prediction was failed.  
In 1982 alignment the planets spread out over 98 degrees \citep{de79}.

It should be noted here that some of the relatively short--term predictions 
of the solar activity which were made   
 based on the hypothesis  of a role of solar system dynamics
 in the  mechanism of solar activity have failed \citep{me91,li01}.
    A reason for this may be
the underlying  physics is not clear. 
 On the other hand,   
the inclinations of the orbital planes  of the planets and   
the  Sun's equator  to the ecliptic  
(or to the invariable plane) seem to be 
important \citep{bl83,jj96,jj03a,ju00}, but  
they  were not taken into account in most of the earlier
investigations. 

\section{Summary and conclusion}
We have showed  the epochs of the  
 violations of the well known G--O rule
 in pairing of sunspot cycles are close to the epochs of the Sun's 
retrograde orbital motion about  the centre of mass of the solar system.
 From this result  easy 
to know well in advance the epochs of violations of the G--O rule.
The G--O rule is expected to be violated   
 by the  Hale cycle  which will include (or end at)  
the year 2169, i.e., only after 
a gap of about eight Hale cycles after the current Hale cycle~11. 
However, the violation of the G--O rule
 by virtue of the difference in the values of the 
peaks of the cycles' pair
--like  the  cycles' 
pair 8,9 near the year 1851--is expected  to be happening near the year 2030, i.e., by  the
 cycles' pair 26,27. 
 We also showed that   
 the solar equatorial rotation rate 
 determined from the  sunspot 
group data during the period 
1879--2004 is correlating to the Sun's orbital torque,
 positively before 1945 and negatively  
after that time. The equatorial rotation has a dominant periodicity 
at  $\sim$ 17 yr.  
These results are well consistent  with the results in the model of 
 the spin--orbit coupling of an oblate Sun by \cite{ju00}, and may 
provide a direct observational support to the hypothesis that 
a role of solar dynamics 
on the internal dynamics of the Sun and in the variations of solar 
activity.

\section*{Acknowledgments}
The author is thankful to  Dr. Ferenc Varadi for  providing  the entire  
 planetary data  used here  and for a fruitful  discussion on the results. 
The author thanks also to Professor Roger K. Ulrich for comments and to the
anonymous referee for useful suggestions.
The author is presently working for the Mt. Wilson Archive Digitization 
project at UCLA, funded by NSF grant ATM--0236682.

\bsp

\label{lastpage}

\clearpage

\begin{thebibliography}{}
\bibitem[Abarbanell \& W\"ohl(1981)]{abrw81}
Abarbanell C., W\"ohl H., 1981, Sol. Phys., 70, 197 
\bibitem[Beer et al.(1990)]{beet90}
Beer J., Blinov A., Bonani G., Hofmann H. J., Finkel R. C., 1990, 
Nature, 347, 164
\bibitem[Bonev, Penev \& Sello(2004)]{boet04}  
Bonev B. P., Penev K. M.,  Sello S. 2004, ApJ,  605, L81
\bibitem[Blizard(1983)]{bl83}
Blizard J. B, 1983, BAAS, 15, 906  
\bibitem[Blizard(1989)]{bl89}
Blizard J. B, 1989, PASP, 101, 890 
\bibitem[Brown(1900)]{brown00}
Brown E. W., 1900, MNRAS, 60(10), 599
\bibitem[Charv\'atov\'a(2000)]{ch00}
Charv\'atov\'a I.: 2000, Ann. Geophys. 18, 399  
\bibitem[DeYoung(1979)]{de79}
DeYoung D. B., 1979, http://www.icr.org/pubs/imp/imp-072.htm
\bibitem[Echer et al.(2004)]{echer04}
Echer E., Rigozo N. R., Nordemann D. J. R., Vieira L. E. A., 2004, Ann. Geophys. 22, 2239
\bibitem[Eddy, Gilman \& Trotter(1976)]{egt76}
Eddy J. A., Gilman P. A., Trotter D. E., 1976, Sol. Phys., 46, 3
\bibitem[Fairbridge \& Shirley(1987)]{fs87}
Fairbridge R. W.,  Shirley J. H., 1987, Sol. Phys., 110, 191
\bibitem[Ferris(1969)]{fe69}
Ferris G. A. J., 1969, J. Brit. Astron. Assoc. 79, 385
\bibitem[Foweler \& Kilsby(2000)]{fow00}
Fowler H. J., Kilsby C. G., 2002, J. Hydrol, 262, 177
 (http://www.staff.ncl.ac.uk/h.j.fowler/histdrought.htm)
\bibitem[Flamsteed(1684)]{fstd1684}
Flamsteed J. 1684, Phil. Trans. 14, 535 
\bibitem[Georgieva et al.(2005)]{geet05}
Georgieva K., Kirov B., Javaraiah J., Krasteva R., 2005, Planet. 
Space Sci., 53, 197
\bibitem[Gnevyshev \& Ohl(1948)]{go48}
Gnevyshev M. N.,  Ohl A. I., 1948, AZh,  25(1), 18
\bibitem[Gokhale(1996)]{gok96}  
Gokhale M. H., 1996, Bull. Astron. Soc. India,  24, 121
\bibitem[Gokhale \& Javaraiah(1995)]{gj95} 
Gokhale M. H., Javaraiah J., 1995, Sol. Phys., 156, 157
\bibitem[Gokhale \& Javaraiah(2002)]{gj02} 
Gokhale M. H., Javaraiah J, (eds),  2002,  The Sun's Rotation. 
Nova Science Publishers, Inc., New York, p. 109 
\bibitem[Gribbin \& Plagemann(1974)]{gp74} 
Gribbin J.,  Plagemann S., 1974,  The Jupiter Effect.  
Walker and CO., New York, p. 261
\bibitem[Hathaway \& Wilson(2005)]{hw05} 
Hathaway D. H., Wilson R. M., 2005, Sol. Phys., 224, 5 
\bibitem[Hathaway, Wilson \& Reichmann(1999)]{het99} 
Hathaway D. H., Wilson R. M.,  Reichmann E. J., 1999, J. Geophys. Res.,
 104, 22375
\bibitem[Hiremath(2002)]{hi02} 
Hiremath K. M., 2002, A\&A, 386, 674
\bibitem[Hiremath \& Mandi(2004)]{hm04} 
Hiremath K. M.,  Mandi P. I.,  2004, NewA, 9, 651
\bibitem[Howard \&  Harvey(1970)]{hh70} 
Howard R., Harvey J., 1970, Sol. Phys., 12, 23
\bibitem[Hoyt \&  Schatten(1997)]{hs97} 
Hoyt D. V.,  Schatten K. H., 1997, The Role of the Sun in Climate Change. 
 Oxford University Press, Inc., New York
\bibitem[Javaraiah(1996)]{jj96} 
Javaraiah J., 1996,  Bull. Astron. Soc. India, 24, 351 
\bibitem[Javaraiah(1998)]{jj98} 
Javaraiah J., 1998,  in Korzennik S. G.,  Wilson A., eds,
 Structure and Dynamics of the Interior of the Sun and Sun-Like Stars,   
ESA SP--418, Noordwijk: ESA, p. 809
\bibitem[Javaraiah(2003a)]{jj03a} 
Javaraiah J., 2003a, Sol. Phys.,  212, 23
\bibitem[Javaraiah(2003b)]{jj03b} 
Javaraiah J., 2003b, A\&A, 401, L9 
\bibitem[Javaraiah \& Gokhale(1995)]{jg95} 
Javaraiah J.,  Gokhale M. H.,  1995, Sol. Phys.,  158, 173 
\bibitem[Javaraiah \& Gokhale(1997)]{jg97} 
Javaraiah J.,  Gokhale M. H., 1997, A\&A, 327, 795 
\bibitem[Javaraiah \& Komm(1999)]{jk99} 
Javaraiah J.,  Komm R. W.,  1999, Sol. Phys., 184, 41 
\bibitem[Javaraiah, Bertello \& Ulrich(2005a)]{jjet05a}
Javaraiah J., Bertello L., Ulrich R. K., 2005a, ApJ, 626, 579 
\bibitem[Javaraiah et al.(2005b)]{jjet05b}
Javaraiah J., Bertello L., Ulrich R. K., 2005b, Sol. Phys. in press 
\bibitem[Jose(1965)]{jo65}
Jose P. D., 1965, AJ,  70, 193
\bibitem[Joselyn et al.(1997)]{jet97} 
Joselyn J., et al., 1997, Eos. Trans. AGU 78, 205, 211
\bibitem[Juckett(2000)]{ju00} 
Juckett D. A., 2000, Sol. Phys.,  191, 201  
\bibitem[Juckett(2003)]{ju03} 
Juckett D. A., 2003, A\&A,  399, 731  
\bibitem[Kambry \& Nishikawa(1990)]{kn90}
Kambry M. A.,  Nishikawa J., 1990, Sol. Phys., 126, 89
\bibitem[Kane(1999)]{kane99}
Kane R. P., 1999, Sol Phys., 189, 217 
\bibitem[Kirov, Georgieva \& Javaraiah(2002)]{kiet02} 
Kirov B., Georgieva K.,  Javaraiah J., 2002, in Wilson A., ed,  
 Solar Variability: 
From Core to Outer Frontiers, ESA SP--506, Noordwijk: ESA, 
p. 149
\bibitem[Komitov \& Bonev(2001)]{kb01} 
Komitov B.,  Bonev B., 2001, ApJ,  554, L119
\bibitem[Krivova, Solanki \& Beer(2002)]{kriet02}
Krivova N. A., Solanki S. K., Beer J., 2002, A\&A, 396, 235
\bibitem[Landscheidt(1999)]{la99} 
Landscheidt T., 1999, Sol. Phys., 150, 359
\bibitem[Li, Yun \& Gu(2001)]{li01} 
Li K. J., Yun H. S.,  Gu X. M., 2001, A\&A, 368, 285
\bibitem[Meeus(1991)]{me91} 
Meeus J., 1991, J. Br. Astron. Assoc. 101, 115  
\bibitem[Mosely(1996)]{mos96}
Mosely J., 1996, Planetarian, 6 (http://www.griffithobs.org/SkyAlignments.html)
\bibitem[M\"orth \& Schlamminger(1979)]{ms79}
M\"orth H. T., Schlamminger L., 1979, in McCormac B. M.,
  Seliga T. A., eds,    
 Solar--Terrestrial Influences on 
Weather and Climate. D. Reidel Publishing Co.,  
Dordrecht, Holland, p. 183
\bibitem[Newton \& Nunn(1951)]{nn51}
Newton H. W., Nunn M. L., 1951, MNRAS, 111, 413
\bibitem[Ossnderijver(2003)]{os03} 
Ossnderijver M., 2003, A\&AR, 11, 287
\bibitem[Ribes \& Nesme-Ribes(1993)]{rnr93}
Ribes J. C., Nesme-Ribes E., 1993, A\&A, 276, 549
\bibitem[Rosner \& Weiss(1992)]{rw92}
Rosner R.,  Weiss N. O., 1992, in Harvey K. L., ed, ASP Conf. Ser. 
Vol. 27,   The Solar Cycle. Astro. Soc. Pac., San Francisco, p. 511 
\bibitem[Rozelot(2001)]{roz01}
Rozelot J. P., 2001, J. Atmos. Solar Terr. Phys, 63, 375
\bibitem[Schove(1955)]{sch55}
Schove D. J., 1955, J. Geophys. Res. 60, 127
\bibitem[Seidelmann(1992)]{sei92}
Seidelmann P. K., 1992,
Explanatory Supplement to The Astronomical Almanac
(Revised edition; Mill Valley, CA: University Science Books)
\bibitem[Sivaraman et al.(2003)]{siet03}
Sivaraman K. R., Sivaraman H., Gupta S. S.,  Howard R. F., 2003, Sol. Phys., 214, 65 
\bibitem[Smythe \& Eddy(1977)]{smy77}
Smythe C. M.,  Eddy J. A., 1977, Nature, 266, 434
\bibitem[Sperber \& Fairbridge(1990)]{sf90}  
Sperber K. R.,  Fairbridge R. W., 1990, Sol. Phys., 127, 379
\bibitem[Standish(1998)]{stand98}
Standish E. M., 1998, JPL Planetary and Lunar Ephemerides, DE405/LE405,
Interoffice Memo. 312.F-98--048, Jet Propulsion Laboratory, Pasadena,
California (Electronic form:
ftp://navigator.jpl.nasa.gov/ephem/export/de405.iom)
\bibitem[Svalgaard, Cliver \& Kamide(2005)]{sval05}
Svalgaard L.,  Cliver E. W.,  Kamide Y., 2005, Geophys. Res. Lett., 32, L01104 
\bibitem[Usoskin, Mursula \& Kovaltsov(2001)]{uset01}  
Usoskin I. G., Mursula K.,  Kovaltsov G. A. 2001, A\&A, 370, L31
\bibitem[Usoskin et al.(2003)]{uset03}  
Usoskin I. G., Mursula K.,  Kovaltsov G. A. 2003, A\&A, 403, 743 
\bibitem[Vaquero, S\'anchez-Bajo \& Gallego(2002)]{vaet02}
Vaquero J. M., S\'anchez-Bajo F.,  Gallego M. C. 2002, Sol. Phys., 207, 219
\bibitem[Wilson(1988)]{wil88} 
Wilson R. M. 1988, Sol. Phys., 117, 269
\bibitem[Wood \& Wood(1965)]{ww65} 
Wood R. M.,   Wood K. D., 1965, Nature, 208, 129 
\bibitem[Zaqarashvili(1997)]{zaq97} 
Zaqarashvili T. V., 1997, ApJ,  487, 930
\end{thebibliography}
\end{document}